\documentclass[sn-mathphys-num]{sn-jnl}

\usepackage{graphicx}%
\usepackage{multirow}%
\usepackage{amsmath,amssymb,amsfonts}%
\usepackage{amsthm}%
\usepackage{mathrsfs}%
\usepackage[title]{appendix}%
\usepackage{xcolor}%
\usepackage{textcomp}%
\usepackage{manyfoot}%
\usepackage{booktabs}%
\usepackage{algorithm}%
\usepackage{algorithmicx}%
\usepackage{algpseudocode}%
\usepackage{listings}
\usepackage{gensymb}%

\theoremstyle{thmstyleone}%
\theoremstyle{thmstyletwo}%

\theoremstyle{thmstylethree}%

\begin{document}

\title[Article Title]{Estimating the impact of light pollution on quantum communication between QEYSSat and Canadian quantum ground station sites}


\author*[1]{\fnm{Mathew} \sur{Yastremski}}\email{mathew.yastremski@ucalgary.ca}

\author*[2]{\fnm{Paul J.} \sur{Godin}}\email{paul.godin@uwaterloo.ca}

\author[2]{\fnm{Nouralhoda} \sur{Bayat}}\email{n3bayat@uwaterloo.ca}
\author[2]{\fnm{Sungeun} \sur{Oh}}\email{paul.rev.oh@uwaterloo.ca}
\author[1]{\fnm{Ziheng} \sur{Chang}}\email{ziheng.chang@ucalgary.ca}
\author[2]{\fnm{Katanya B.} \sur{Kuntz}}\email{katanya.kuntz@uwaterloo.ca}
\author[1]{\fnm{Daniel} \sur{Oblak}}\email{doblak@ucalgary.ca}
\author*[2,3]{\fnm{Thomas} \sur{Jennewein}}\email{tjennewe@uwaterloo.ca}

\affil[1]{\orgdiv{Institute for Quantum Science and Technology, Department of Physics and Astronomy}, \orgname{University of Calgary}, \orgaddress{\street{2500 University Dr NW}, \city{Calgary}, \postcode{T2N 1N4}, \state{Alberta}, \country{Canada}}}

\affil[2]{\orgdiv{Institute for Quantum Computing, Department of Physics and Astronomy}, \orgname{University of Waterloo}, \orgaddress{\street{200 University Ave W}, \city{Waterloo}, \postcode{N2L 3G1}, \state{Ontario}, \country{Canada}}}

\affil[3]{\orgdiv{Department of Physics}, \orgname{Simon Fraser University}, \city{Burnaby}, \country{Canada}.}

\abstract{Satellite to ground quantum communication typically operates at night to reduce background signals, however it remains susceptible to noise from light pollution of the night sky. In this study we compare several methodologies for determining whether a Quantum Ground Station (QGS) site is viable for exchanging quantum signals with the upcoming Quantum Encryption and Science Satellite (QEYSSat) mission. We conducted ground site characterization studies at three locations in Canada: Waterloo, Ontario, Calgary, Alberta, and Priddis, Alberta. Using different methods we estimate the background counts expected to leak into the satellite-ground quantum channel, and determined whether the noise levels could prevent a quantum key transfer. We also investigate how satellite data recorded from the Visible Infrared Imaging Radiometer Suite (VIIRS) can help estimate  conditions of a particular site, and find reasonable agreement with the locally recorded data. Our results indicate that the Waterloo, Calgary, and Priddis QGS sites should allow both quantum uplinks and downlinks with QEYSSat, despite their proximity to  urban centres. Furthermore, our approach allows the use of satellite borne instrument data (VIIRS) to remotely and efficiently determine the potential of a ground site.}




\maketitle

\section{Introduction}\label{sec1}

As quantum communication capabilities advance, the need for larger quantum-based networks has become evident. While optical fiber is very well established, the length of transmission is limited to a few hundred kilometres due to photon loss \cite{korzh2015provably, inagaki2013entanglement}. Quantum repeaters will eventually be able to surpass the distance limitations of ground-based quantum channels \cite{avis2023requirements,briegel1998quantum}, however this technology is yet to be demonstrated. One established solution for continental scale quantum communication is to use  satellites and implement ground-to-space quantum links \cite{liao2017satellite,chen2021integrated,Jennewein2014}. 

The Canadian Quantum Encryption and Science Satellite (QEYSSat) mission is intended to be a scientific platform and demonstrator with the primary mission objective to develop and demonstrate the capability to distribute highly secure encryption keys from a Quantum Ground Station (QGS) to a satellite. This mission aims to advance the technical readiness of quantum communication using ground-to-space links, and study the science related to such a quantum channel. One mission goal is to demonstrate ground-to-space Quantum Key Distribution (QKD), which is a secure method of establishing encryption keys between two distant parties using single photons. While traditional encryption methods rely on mathematical complexity to protect information, QKD relies on the laws of quantum mechanics to protect data \cite{scarani2009security,bennett1984proceedings,ekert1991prl}. 

The primary functionality of the QEYSSat payload is a receiver for a quantum uplink using a 25\:cm aperture telescope. One significant benefit of using an uplink connection is the capability to use different types of photon emitters at the QGS. A recent addition to the QEYSSat payload is a quantum source module that will demonstrate a QKD downlink channel using a novel reference frame independent protocol \cite{QeyssatRFI}. A crucial preparation stage for this mission is to characterize the light pollution conditions at various QGS locations to ensure successful quantum links to QEYSSat.

With the recent push for long-distance free-space QKD, there has been rising interest for the installation of QGSs, especially near urban environments where most potential users of a quantum network are located. This interest means there is a necessity to consider how to classify whether a location is suitable for a QGS. While many factors need to be considered, such as proximity to existing fibre optic infrastructure and light pollution conditions, a primary determinant for a possible QGS site is whether a quantum link is viable (i.e. can the quantum signal be distinguished from background noise). One approach is to consider the quantum bit error rate (QBER) for a QKD link. The QBER is generally expressed as a fraction of incorrectly detected bits in a sequence divided by the total number of expected detections \cite{vallone2015experimental,ribordy2000fast, ricklin_atmospheric_2006,liao2017satellite}. Any noise or loss in the channel will increase the QBER until it's no longer possible to establish a secure key. There are other protocols than QKD, such as secure time transfer, that could still make use of a higher loss quantum link, and this is an active area of research.

In light of increased urbanisation and energy consumption globally, Kyba \textit{et. al} studied the increase of artificial light present on Earth at night \cite{Kybae1701528}. The first-ever calibrated satellite radiometer designed for night lights was used to show that from 2012 to 2016, Earth’s artificially lit outdoor area grew by 2.2\% per year, with a total radiance growth of 1.8\% per year. This measured and increasing light pollution may complicate the detection of quantum signals used for satellite quantum communication \cite{er-long_background_2005,ricklin_atmospheric_2006}.

In this study, we examine the expected impact of light pollution on free-space satellite quantum communication within QEYSSat's wavelength range of interest (750\:nm -- 850\:nm) at three Canadian locations: University of Waterloo campus (QGS-UW), the University of Calgary campus (QGS-UC), and Rothney Astrophysical Observatory (QGS-RAO).  QGS-UW and QGS-UC are located in close proximity to their respective urban centres, thus, yielding relatively high light pollution levels. However, the proximity of these two QGS to a laboratory and potential city-wide optical networks adds a  benefit for the utility of a QKD link. The third location QGS-RAO, which resides in a dark sky area 40\:km from QGS-UC, is also used to compare a rural location to the two urban locations.

We employed two analytical methodologies to form an experimental guide for selecting candidate locations for future QGS locations. The first method presented is ground-based observation of the night-sky designed to find an upper bound on the expected background light levels due to local light pollution conditions. The second method used NASA-NOAA Suomi NPP satellite data taken by the Visible Infrared Imaging Radiometer Suite (VIIRS) Day and Night band (DNB) to estimate local light-pollution conditions of a site. The DNB data used in this study is ultra-sensitive in low-light conditions, allowing observation of nighttime lights with better spatial and temporal resolutions compared to previously provided nighttime lights data by the Defence Meteorological Satellite Program (DMSP). The DNB used by VIIRS is a night-time sensor that captures light emission in a broad spectral range from 500\:nm to 900\:nm \cite{ROMAN2018113, VIIRS_Technical_report}. Determining the correct approach to harmonize the broadband VIIRS data to our local measurements is essential to analyse the case of QEYSSat being used as a receiver, i.e., the uplink scenario. Further information on VIIRS and the data it supplies the user guide is available at \cite{VIIRS_Technical_report}.

\section{Methods}\label{sec2}

The main source of noise photons we consider here is the local light pollution at a QGS site, within targeted quantum communication wavelength ranges \cite{er-long_background_2005, ricklin_atmospheric_2006, ricklin_bit_2003}. Both the QGS and the quantum payload aboard the satellite operate as transceiver units, and we consider both the downlink and uplink operation of QEYSSat \cite{bourgoin2013comprehensive}. The definition of `downlink' is when the satellite operates as the transmitter (i.e. the sender in QKD, `Alice'), and the QGS functions the receiver (i.e. `Bob'). While in the `uplink' configuration, the QGS serves as the transmitter and the satellite operates as the receiver.

\begin{figure}
\centering
\includegraphics[width = \linewidth]{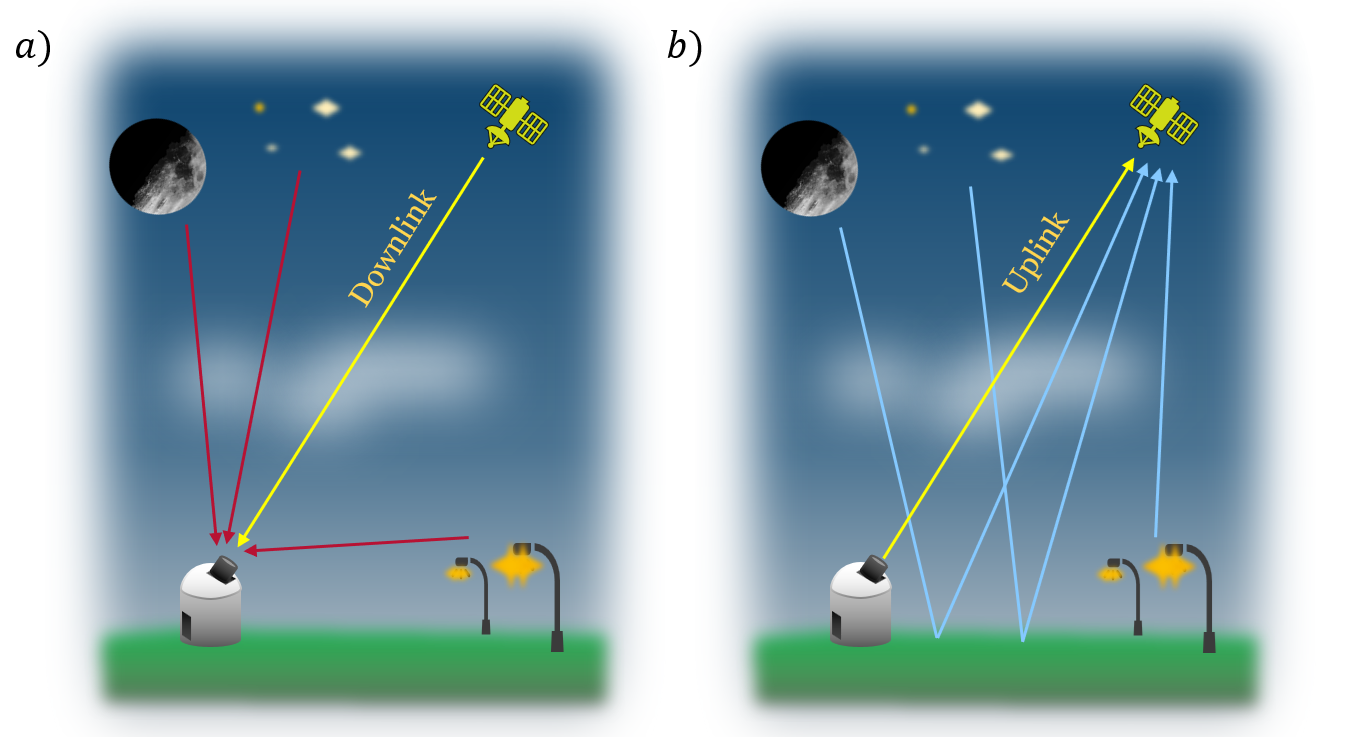}
\footnotesize
\caption{(a) Shows the downlink scenario where the QGS receiver is collecting light from the satellite, as well as scattered moonlight, starlight and reflected urban light pollution. (b) shows the uplink scenario where the satellite receiver is exposed to reflected moonlight, starlight and urban scattered light pollution.
} \label{fig:updown}
\end{figure}

During a downlink, the QGS receiver will pick up background signals from various natural light sources (moon, stars), as well as anthropogenic light scattered in the sky, see Figure~\ref{fig:updown}a. When the satellite receiver is looking down from its orbit towards the QGS during uplink, it will receive the QGS quantum signals in addition to any stray light due to the surrounding street lights,  buildings, and other sources of anthropogenic illumination within the receiver's field-of-view \cite{er-long_background_2005}, see Figure~\ref{fig:updown}b. 

While selecting a remote or rural location may alleviate some of the challenges of anthropogenic light pollution, practical constraints also influence the QGS site selection. Most current QGSs require proximity to laboratories that house quantum sources and other essential instruments. Alternatively, the quantum signals can be relayed from the QGS to a suitable facility over fibre optic links, however, with at the cost of further signal attenuation \cite{chen2021integrated, liao2017satellite}. If we consider the experimental needs of the QEYSSat Science Team during mission operations, an on-campus QGS in close proximity to the relative laboratories would be ideal due to its accessibility and security.

\subsection{Light Pollution for Quantum Downlink}

This section details methods used in analysing satellite downlink light pollution data. When considering downlink satellite communication, the QGS is receiving a quantum signal from an in-orbit satellite,  so any light source that is detectable in the sky between the QGS and the passing satellite must be considered. The QGS-UW and QGS-UC/QGS-RAO teams used two different measurement methods to characterize background noise for a downlink scenario. 

At QGS-UW, we used a highly-portable measurement system consisting of a 0.22 numerical aperture multi-mode fibre that is coupled directly into an Excelitas SPCM-AQRH series single photon detector (SPD). This setup allowed for direct measurement of the photon flux in a given solid angle. During data collection, the setup was covered by an opaque tarpaulin on both ends to prevent extraneous light from entering and to increase measurement accuracy. The fibre cage system was oriented in eight distinct azimuth directions at three elevation angles that correspond to the satellite's orbital path, as listed in Table \ref{tab:Locations}. Measurements were taken over 5 second intervals at each position, and normalised to photons/second. Optical filters were used to take measurements at specific wavelengths: 780\:nm, 790\:nm and 850\:nm, which are listed in Table \ref{tab:filter}.

\begin{table}[h]
\caption{The measurement locations and directions of the three data sets collected at QGS-UW, QGS-UC, and QGS-RAO. *indicates altitudes where measurements were only taken in the four cardinal directions.}\label{tab:Locations}%
\begin{tabular}{@{}lcccccc@{}}
\toprule
 & Detector  & Elevation & Location & \# of & Measurement  & Measurement  \\
 &FOV & (m) & & detectors& azimuth &altitude\\
\midrule
QGS-UW  & 0.010$\degree$   & 329  & 43$\degree$   28' 43.99"N  & 1 & $\pm45\degree$, $\pm90\degree$, & $45\degree$, $65\degree$,  \\
& & & 80$\degree$ 33' 17.77"W & &$\pm135\degree$, $180\degree$ &  $85\degree$ \\
\midrule
QGS-UC  & 0.008$\degree$   & 1109  & 51$\degree$  4' 45.84"N & 4 & $\pm45\degree$, $\pm90\degree$,& $30\degree$, $45\degree$,\\
& & & 114$\degree$ 8'   29.00"W &&$\pm135\degree$, $180\degree$ &  $60\degree$, $85\degree$\footnotemark[1] \\
\midrule
QGS-RAO & 0.008$\degree$   & 1109  & 51$\degree$ 50$\degree$  52' 4.94"N & 4 & $\pm45\degree$, $\pm90\degree$,& $30\degree$, $45\degree$,\\
& & & 114$\degree$ 17'   28.11"W &&$\pm135\degree$, $180\degree$ &  $60\degree$, $85\degree$\footnotemark[1] \\
\botrule
\end{tabular}
\footnotetext[1]{Altitudes where measurements were only taken in the four cardinal directions.}
\end{table}

At QGS-UC and QGS-RAO, a 280\:mm-aperture computer-controlled Cassegrain telescope was used to characterize the background light in eight distinct azimuth directions at four elevation angles, as listed in Table \ref{tab:Locations}. Incoming light collected by the telescope passed through a bandpass filter before being coupled into $105\, \mu m$ multi-mode fibres and sent to Excelitas-SPCM-NIR SPDs. Three different 10\:nm-FWHM bandpass filters with central wavelengths at 750\:nm, 800\:nm, and 850\:nm were used during data collection (see Table \ref{tab:filter}). These NIR (near-infrared) wavelengths are potential candidates for satellite-based QKD using silicon-based SPDs, which are commercially available and work at room-temperatures.

\begin{table}[h]
\caption{Information on the filters and fibres used for data collection. All filters and fibres used were off-the-shelf components.}\label{tab:filter}%
\begin{tabular}{@{}lcccc@{}}
\toprule
 & Filters used  & Filter Bandwidth & Type of Fibre & NA of Fibre \\
 &(nm) & (nm) & & \\
\midrule
 
 QGS-UW  & 780, 790, 850   & 10  & 105\:$\mu$m multi-mode  & 0.22 \\
\midrule
 
  QGS-UC/  &    &  &  & \\
 QGS-RAO  & 750, 800, 850   & 10  & 105\:$\mu$m multi-mode & 0.22\\

\botrule
\end{tabular}

\end{table}


\subsection{Light Pollution for Quantum Uplink}

We employed two different methods to model the background noise detected by the QEYSSat transceiver when pointing down at the QGS. The QGS-UW team used the same fibre-detector cage setup from the downlink measurements, pointing it from the roof toward the ground to collect background light from the area surrounding the QGS-UW site. The photon counts received were used to determine the radiance of the ground surrounding the QGS, which can then be extrapolated to photons arriving at QEYSSat. We assume that any atmospheric absorption effects are negligible over this short distance and can be ignored $(e_{atm}=1)$.

\begin{figure}
\centering
\includegraphics[width = \linewidth]{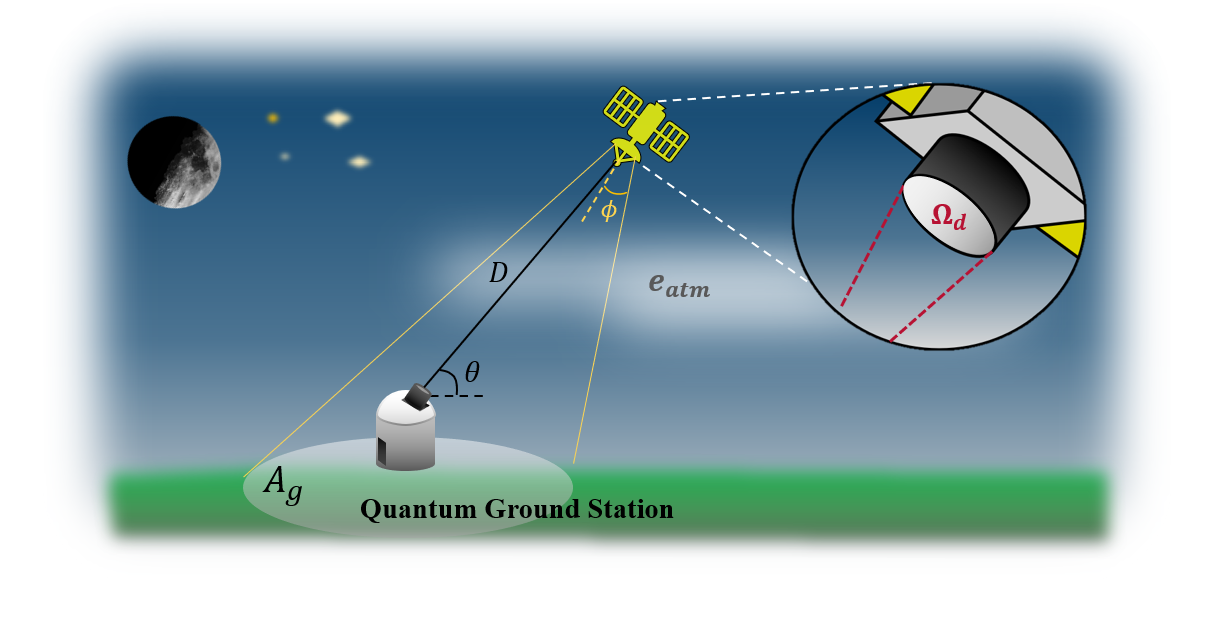}
\footnotesize
\caption{An illustration of the geometry of a passing satellite when pointing at a QGS during the uplink scenario. $A_g$ is the visible area the satellite sees, $D$ is the distance between the QGS and satellite, $\theta$ is the altitude angle of the satellite above horizon, $\phi$ is the half angle of acceptance of the satellite, $e_{atm}$ is the atmospheric absorption coefficient, and $\Omega_d$ is the solid angle of the detector as seen from the ground. During the downlink scenario $A_g$ and $\Omega_d$ are interchanged. } \label{fig:Parameters}
\end{figure}

To convert the collected data between photon rate and radiance we used
\begin{equation} \label{eq:photoncount}
    N \ = \ e_{atm}A_{g}\Omega_{d}\frac{L}{E_{\lambda}}, 
\end{equation}
where $N$ is the number of photons, $e_{atm}$ is the absorption through the atmosphere, $L$ is the radiance of the ground, and $E_{\lambda}$ is the energy of a single photon. $A_{g}= \pi D^2 \phi^2$ is the area of the emitting source seen by the detector, with $\phi$ being the half-angle of acceptance of the detector, and $D$ the distance between the source and detector. $\Omega_{d} = \pi(r_{sat}/D)^{2}$ is the solid angle of the detector as seen from the ground, where $r_{sat}$ is the radius of the detector. 
Making the appropriate substitutions yields:
\begin{equation}\label{eq:raidance}
    L = E_{\lambda}\frac{N}{(\pi\phi r_{sat})^{2}e_{atm}}.
\end{equation}

For a satellite looking down at the same ground surface, we can expand Equation~\ref{eq:photoncount} to account for the satellite altitude angle above the horizon, which impacts both the area of the ground seen and the atmospheric absorption. The satellite will see an area of
\begin{equation} \label{eq:Sat_ground}
    A_{g} \ = \ \pi(\tan(\phi)D)^{2}\csc(\theta),
\end{equation}
where $\phi$ is again the half-angle of acceptance of the satellite receiver, and $\theta$ is the altitude angle of the satellite. These parameters are also defined in Figure \ref{fig:Parameters}. From \cite{bourgoin2013comprehensive}, the atmospheric absorption can be approximated by
\begin{equation} \label{eq:atmo_abs}
   e_\mathrm{atm} \ = \ 10^{-0.32\csc(\theta)}.
\end{equation}

Substituting Equations~\ref{eq:Sat_ground} and \ref{eq:atmo_abs} into Equation~\ref{eq:photoncount}, the single photon counts reaching the satellite as a function of altitude angle can now be expressed in terms of photon rate measured on the roof,
\begin{align}\label{eq:sat_counts}
    N_\mathrm{sat}(\theta) &= 10^{-0.32\csc(\theta)}\tan(\phi)^{2}\csc(\theta)r_\mathrm{sat}^{2}\frac{N_{r}}{(\alpha r_{f})^{2}},
\end{align}
where $N_\mathrm{sat}(\theta)$ is the number of photons reaching the satellite, $N_{r}$ is the number of photons measured by the detector on the roof looking at the ground, $r_{f}$ is the radius of the fibre core on the rooftop detector, $\alpha$ is the half-angle of the fibre, and $r_\mathrm{sat}$ is the radius of the satellite receiver.

\begin{figure}[h]
     \centering
     \includegraphics[width = \linewidth]{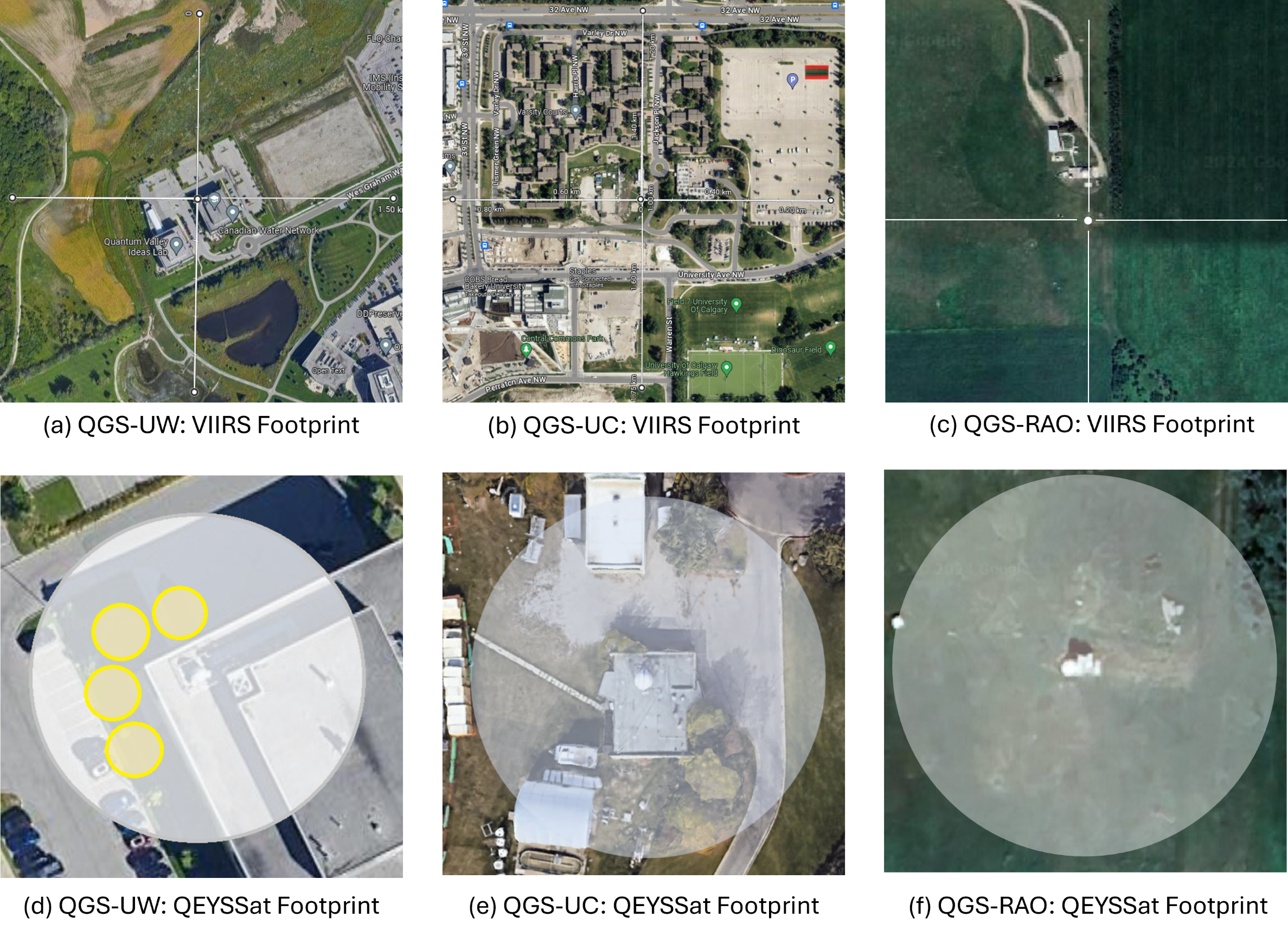}

        \caption{Satellite images of QGS-UW, QGS-UC, and QGS-RAO taken from Google Maps. Images (a), (b), and (c) show the 500\:m $\times$ 500\:m area of a single VIIRS pixel centred on each ground station. The light grey circle in images (d), (e), and (f) identifies the approximate footprint the QEYSSat detectors will see on the ground at zenith during an uplink scenario, assuming an orbit of 550~km altitude. The yellow circles in (d) identify the areas measured at QGS-UW using the experimental fibre measurement set-up on the corner of the roof pointing down at the ground.}
        \label{fig:google_views}
\end{figure}

Figure~\ref{fig:google_views} shows satellite images of QGS-UW, QGS-UC and QGS-RAO, highlighting the footprint QEYSSat will see at zenith (assuming a 550\:km low-earth-orbit) as light grey circles in figures~\ref{fig:google_views}(d), (e) and (f). Whereas the measurement areas of ground radiance made from the rooftop at QGS-UW are shown as yellow circles in figure~\ref{fig:google_views}(d). These four ground measurements were averaged together and used to estimate the uplink light pollution that could reach QEYSSat.

The second method used was to process ground radiance values taken by the NASA-NOAA Suomi NPP satellite VIIRS Instrument \cite{VIIRS_Technical_report}. The DNB used by VIIRS is a night-time sensor that captures light emission in the spectral range of 500\:nm to 900\:nm at a ground resolution of 500\:m $\times$ 500\:m per pixel \cite{VIIRS_Technical_report, ROMAN2018113}. VIIRS VPN46A1 data product collected during the same nights as described in Table~\ref{tab:Conditions} were used. Geographic coordinates were used to determine which pixel in the VIIRS images corresponded to the QGSs locations, and the radiance value of each pixel was extracted by use of QGIS software. However, the DNB ground radiance of the pixels needed to be scaled to account for the differences in ground footprints between QEYSSat and VIIRS, as well as the difference in spectral bandwidths.

\begin{figure}[h]
\centering
\includegraphics[width = \linewidth]{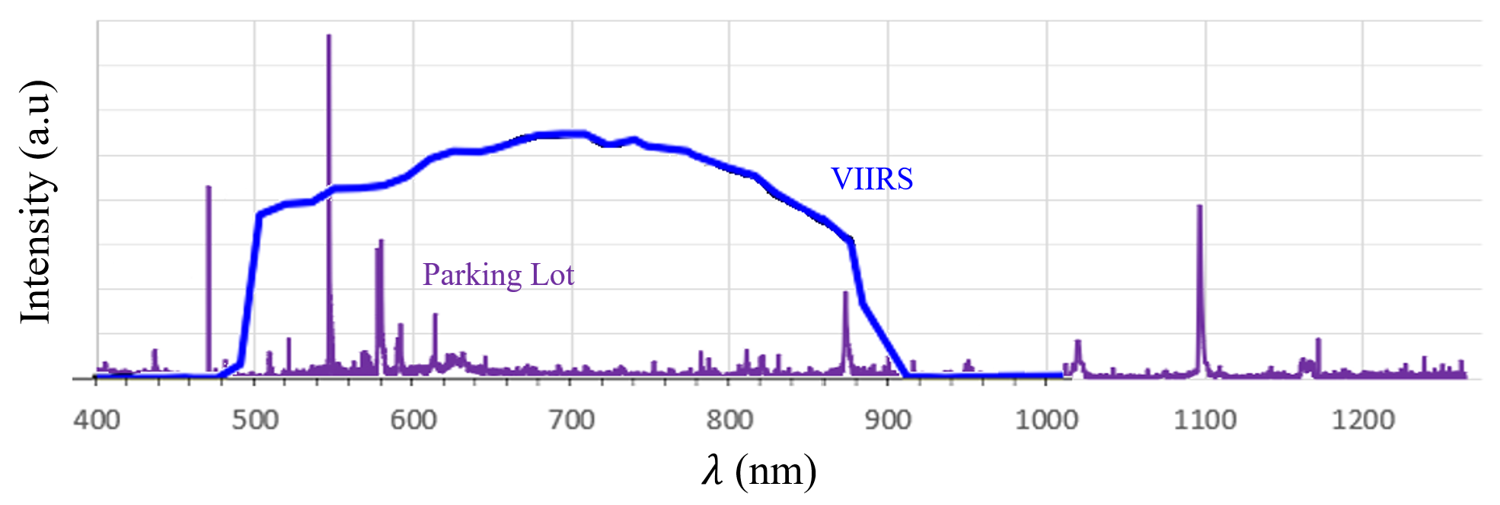}
\footnotesize
\caption{The sensitivity spectrum of the DNB sensor used by VIIRS (in blue) \cite{wang2017viirs}. The dark subtracted spectrum of the area surrounding QGS-UW taken from the rooftop location of the ground station looking down at the parking lot (in purple).} \label{fig:VIIRS}
\end{figure}

To scale the VIIRS spectral range, we characterized the spectral emissions of light sources around QGS-UW with an OceanOptics spectrometer, as shown in Figure~\ref{fig:VIIRS}. The integrated spectral bandstrength in regions of interest was divided against the 500\:nm to 900\:nm integrated spectral bandstrength, to obtain the following spectral scaling factors: $0.7\%\pm0.2\%$ for 780\:nm, $0.7\%\pm0.2\%$ for 790\:nm, and $0.5\%\pm0.2\%$ for 850\:nm. 

To scale the footprints of VIIRS to match QEYSSat, the ratio of illuminated areas $A_\mathrm{ill}$ (buildings/roads/parking lots) to dark areas $A_\mathrm{dark}$(fields/parks) was used to scale up the VIIRS value $L_\mathrm{VIIRS}$ as if the entire field of view was illuminated, assuming dark areas have a radiance of $L_\mathrm{dark}=0$:
\begin{align}
        L_\mathrm{VIIRS} \ &= \ L_\mathrm{ill}A_\mathrm{ill} + L_\mathrm{dark}A_\mathrm{dark} \\
  L_\mathrm{ill} \ &= \  \frac{L_\mathrm{VIIRS}}{A_\mathrm{ill}}
\end{align}
The illuminated area fractions are estimated from Figures~\ref{fig:google_views}a, \ref{fig:google_views}b, and \ref{fig:google_views}c as approximately 1/3, 3/4, and 1/10 $\pm 20\%$ respectively. Next the values are scaled down to match the fraction of the QEYSSat field of view that is illuminated in Figures~\ref{fig:google_views}d, \ref{fig:google_views}e, and \ref{fig:google_views}f as approximately 3/4, 1/2, and 1/10 $\pm 20\%$ respectively. The total uncertainty for the VIIRS scaling is estimated to be $\pm40\%$. 

\section{Results}
\subsection{Quantum Downlink}
Measurements were recorded at QGS-UW on two separate nights. The first was November 23, 2022 which was a new moon and dry ground conditions. The second was February 7, 2023, which was a full moon and snow-covered ground conditions. These conditions were thought to be considered the extreme cases of light pollution that would be experienced at the QGS location. We have summarized the measurement conditions in Table \ref{tab:Conditions}.

\begin{figure}
\centering
\includegraphics[width = \linewidth]{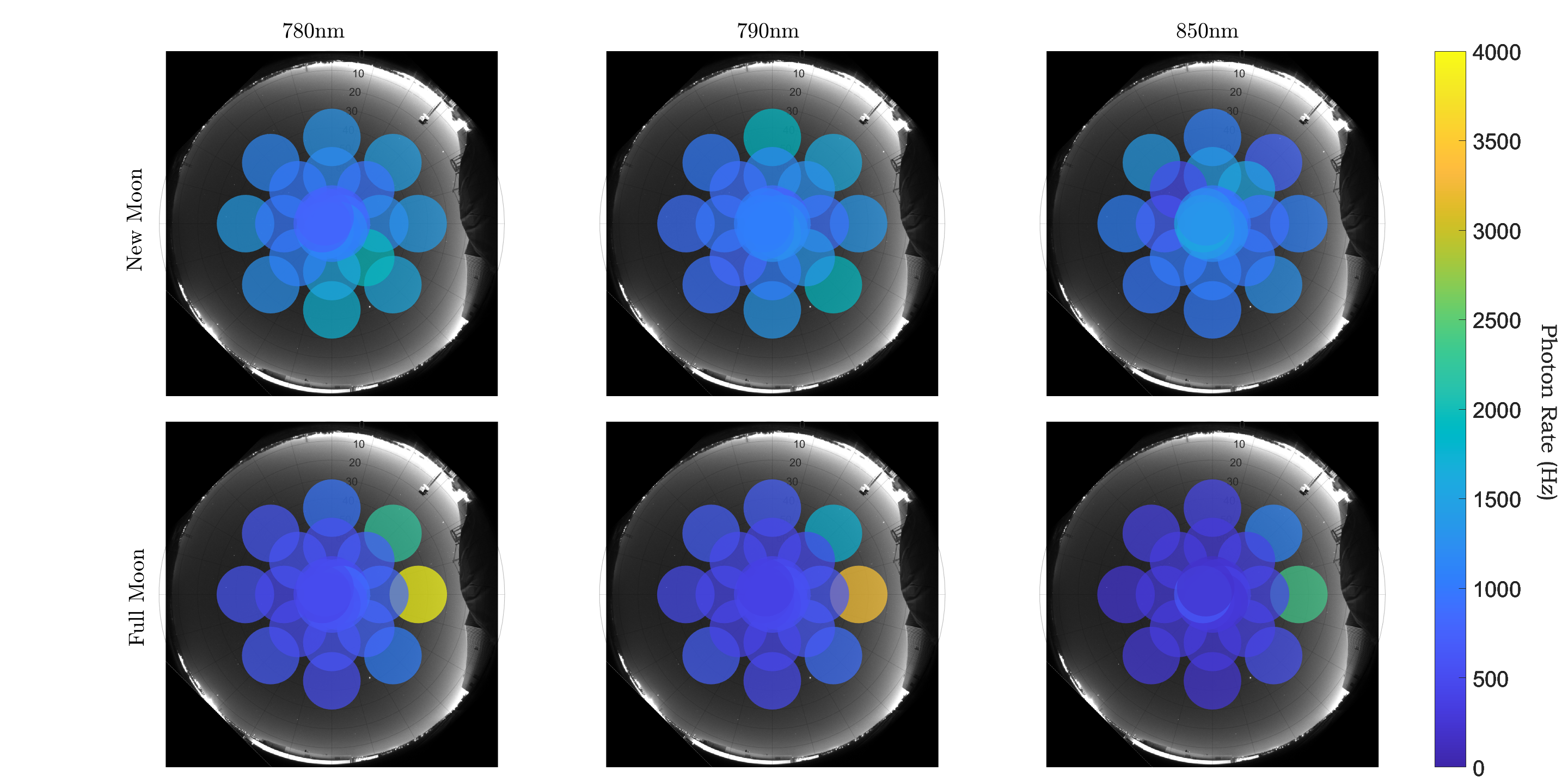}
\footnotesize
\caption{The downlink photon count rates projected at different wavelength ranges at QGS-UW. Data points are scaled in size to match the solid angles of where the data was measured. The top row of images correspond to new moon and dry ground conditions taken on November 23, 2022; while the bottom row corresponds to full moon and snow covered ground taken on February 7, 2023.} \label{fig:Downlink}
\end{figure}

The top row of Figure \ref{fig:Downlink} displays the measurement outcomes across three distinct wavelength bands captured during a new moon. Meanwhile, the bottom row data corresponds to the same wavelength bands, but observed on a night with a full moon and snow-covered ground. The data points are superimposed on an all-sky visible image taken at QGS during their corresponding nights. On average counts are typically between 300-1000 Hz regardless of lunar conditions; but near the vicinity of the full moon, counts rose up to just under 4000 Hz.

\begin{table}[h]

\caption{The moon phases, measurement times, weather conditions, and VIIRS radiance values of each data collection period at QGS-UW, QGS-UC, and QGS-RAO. Full moon measurements were taken while the moon was present.}\label{tab:Conditions}%

\begin{tabular}{@{}lcccccc@{}}
\toprule
  & Measurement & Temp. & Weather & Moonrise & Moon  & VIIRS  \\
 & Time (UTC)& ($\degree$C) & &(UTC) & Illumination&(nW/cm$^2$/sr)\\
\midrule
QGS-UW \\
Nov 24, 2022& 0:00       & 1  & Clear  & 0:04 & 0.1\%   & 37.7  \\
Feb 7, 2023  & 1:00     & -4  & Clear & -0:26    & 99.6\%   & 53.5      \\
\midrule
QGS-UC \\
Mar 20, 2023  & 3:40     & -4  & Clear & 2:43 & 1.3\% & 105.3\\
Feb 8, 2023  & 6:15    & -12  & Clear & 6:06    & 96\%   & 104.7      \\
\midrule
QGS-RAO \\
July 4, 2024    & 7:00   & 12  & Clear & 9:32 & 1.8\% & 2.9  \\
June 24, 2024   & 6:30   & 15  & Clear & 6:02 & 96.3\% & 3.4    \\
\botrule
\end{tabular}

\end{table}

\begin{figure}
\centering
\includegraphics[width = \linewidth]{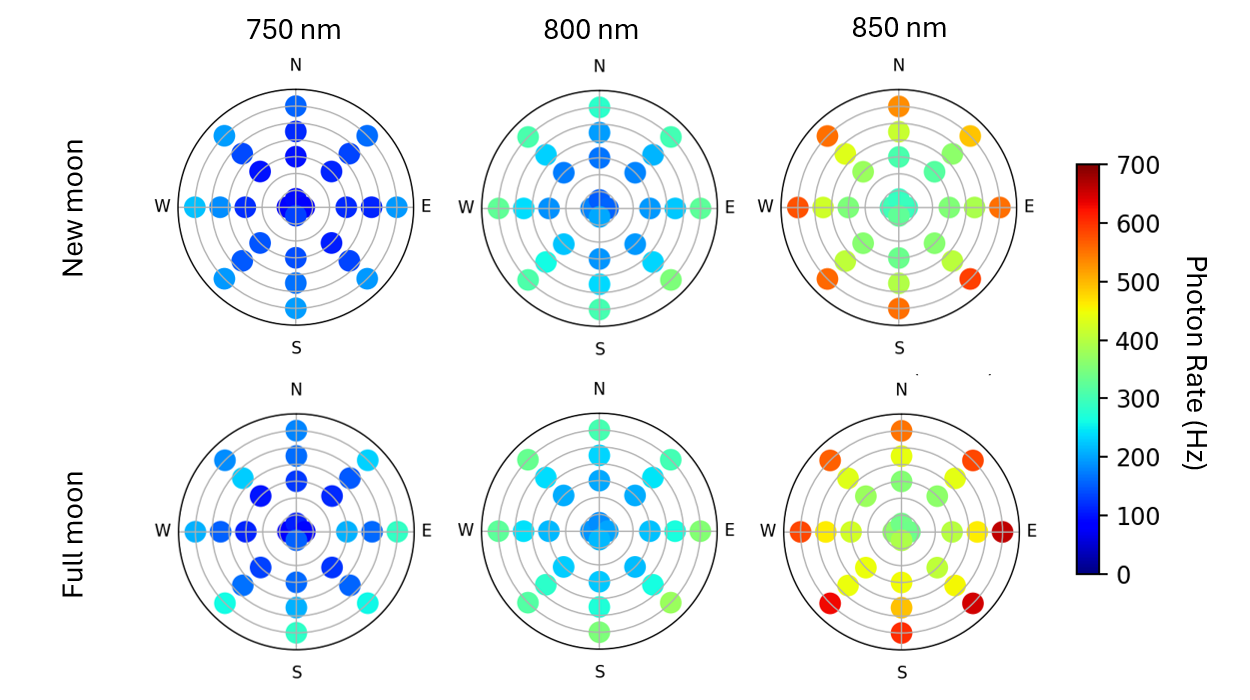}
\footnotesize
\caption{Comparison of new moon and full moon light pollution values in downlink photon rate (Hz) for the three filters used in the telescope of QGS-UC. The filters used have a bandwidth of 10\:nm.}
\label{fig:UofC8Panel}
\end{figure}

\begin{figure}
\centering
\includegraphics[width = \linewidth]{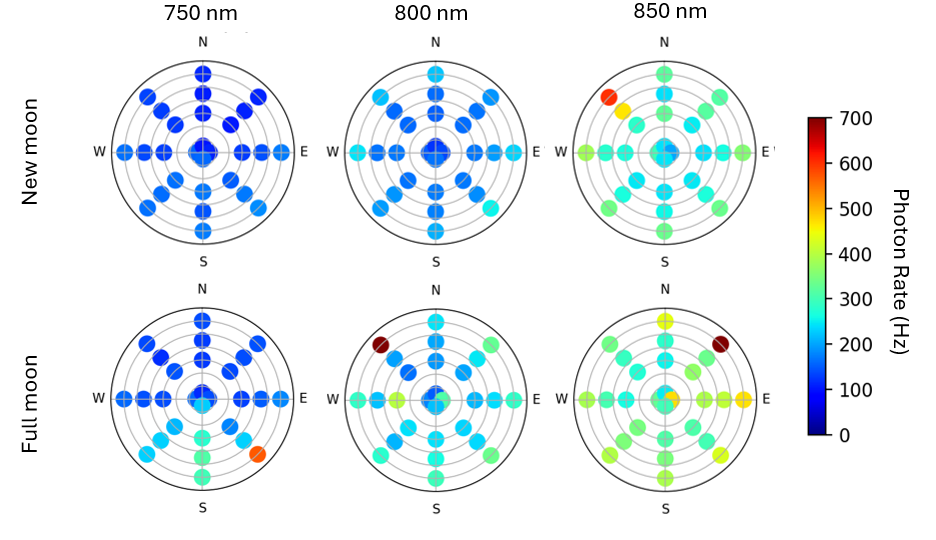}
\footnotesize
\caption{Comparison of new moon and full moon light pollution values in downlink photon rate (Hz) for the three filters used in the telescope of QGS-RAO. The filters used have a bandwidth of 10\:nm.}
\label{fig:RAO8Panel}
\end{figure}

Data from QGS-UC was collected at 28 points in the sky, the directions are shown in Table \ref{tab:Locations}. The altitude/azimuthal positions were chosen based on the observations that most secure keys from satellite-based QKD experiments are generated within this altitude range. At each point of measurement, 30 seconds of data is collected before being normalised to photon rate. The downlink measurements presented in Figure~\ref{fig:UofC8Panel} were collected under both full moon (Feb 8th, 2023) and new moon (Mar 20, 2023) conditions to bound the influences of moonlight at QGS-UC. The results are normalized and presented in photon rate (Hz), allowing for cross-comparison with results from QGS-UW and QGS-RAO.

The photon rates of interest are presented in Figure~\ref{fig:UofC8Panel}. The figure shows a 3d sky map projected onto a polar graph. The four cardinal directions are labelled and each data point moving inward represents an angle of altitude. The outermost ring of points shows 30$\degree$ altitude, while the innermost four points are 85$\degree$ altitude. With the remaining two rows of data points being 45$\degree$ and 60$\degree$. The measurements shown on the graphs in Figure~\ref{fig:UofC8Panel} have been normalised to include detector efficiencies and total efficiency of the telescope's optical components.  Uncertainty in the measurements was propagated statistically while normalising the data.

Data from QGS-RAO was collected in the same manner as QGS-UC. The measurement parameters for QGS-RAO are shown in Table \ref{tab:Locations} where at each point of measurement, 30\:s of data is collected before being normalised to photon rate. The downlink measurements presented in Figure~\ref{fig:RAO8Panel} were collected under both a full moon (June 24, 2024) and new moon (July 4, 2024) conditions to bound the influences of moonlight at QGS-RAO. The results are normalized and presented in photon rate (Hz), allowing for cross-comparison with results from QGS-UW and QGS-UC.

\subsection{Quantum Uplink}
\begin{figure}[h!]
\centering
\includegraphics[width =\linewidth]{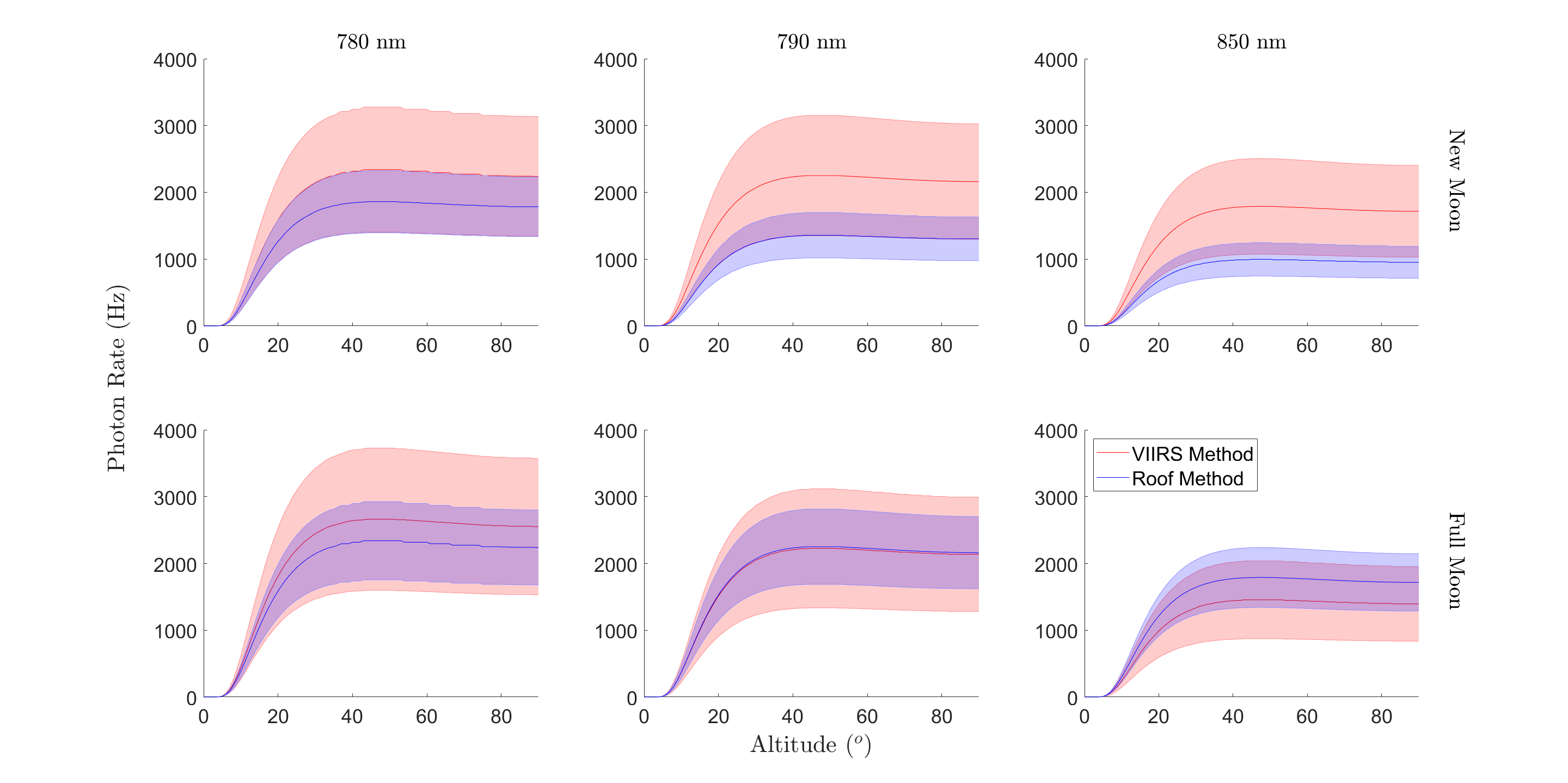}
\footnotesize
\caption{Estimated photon counts reaching QEYSSat for an uplink configuration for QGS-UW. Values derived from the VIIRS method are in red, values derived from the rooftop method are in blue. The rooftop method uncertainty comes from varying the amount of the QEYSSat footprint we consider illuminated by the parking lot (50\%-100\%, centre line at 75\%). The VIIRS uncertainty is from 40\% variation seen in the parking lot spectrum and field of view scaling.} \label{fig:uplink_sum}
\end{figure}

Radiance values extracted from the VIIRS-DNB VPN46A1 data products, at the pixel coordinates associated with the physical locations of Waterloo, Calgary, and Rothney can be used in Equation~\ref{eq:sat_counts} as another estimation of background noise reaching QEYSSat. Figure~\ref{fig:uplink_sum} shows the expected number of background photons reaching QEYSSat from the ground surrounding the QGS-UW. At QGS-UW, the uplink photon rate shown in Figure~\ref{fig:uplink_sum} also reaches the order of 2000\:Hz when using the VIIRS-DNB data, as it shows a radiance of 53.5 nW/cm$^2$/sr. Additionally, the rooftop measurements taken at QGS-UW fall within the anticipated error of the processed VIIRS-DNB data with the 850\:nm new moon measurement being the lowest of the data sets. 

\begin{figure}[h!]
\centering
\includegraphics[width = \linewidth]{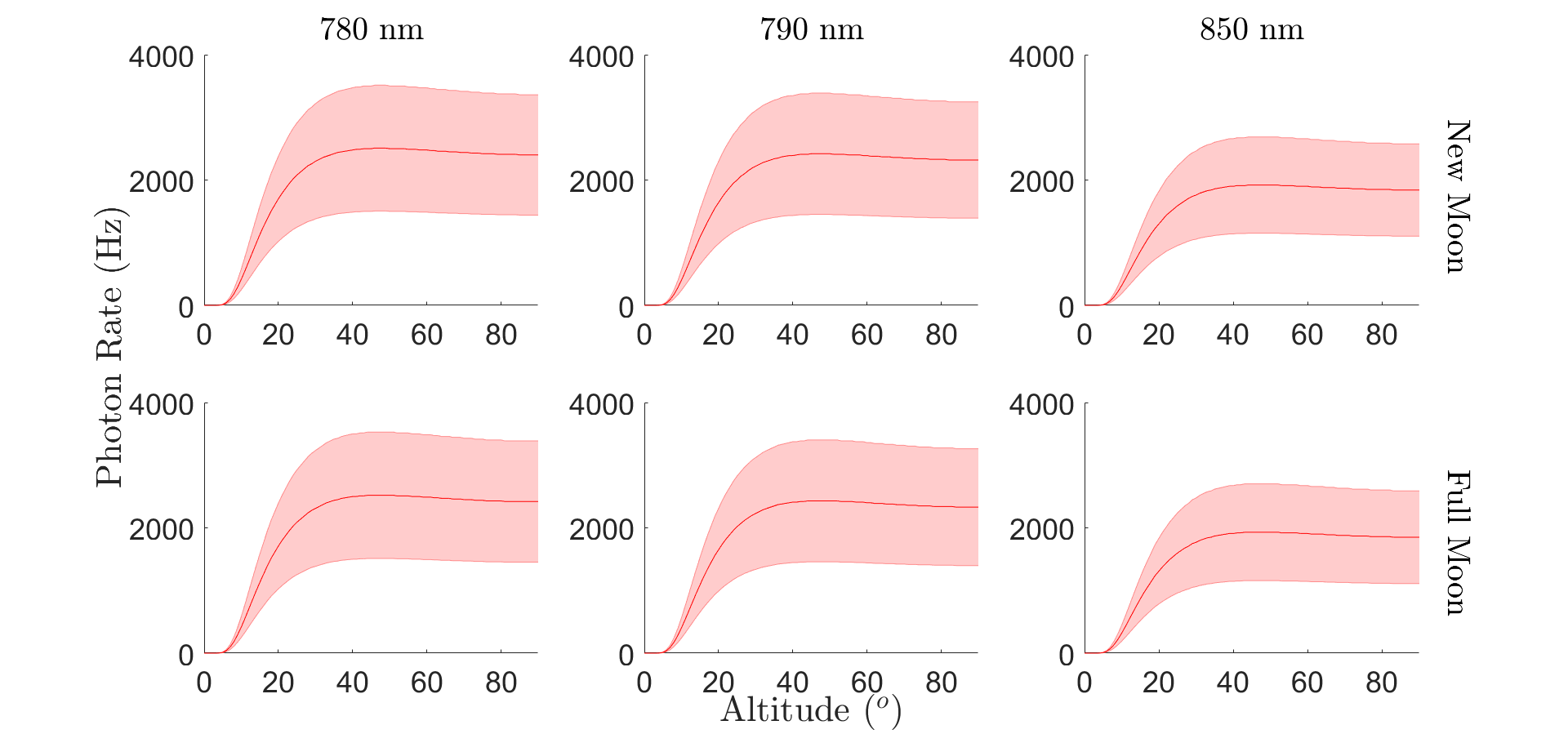}
\footnotesize
\caption{Expected photon counts reaching QEYSSat for an uplink configuration from the QGS-UC using the VIIRS method. The VIIRS uncertainty is from 40\% variation seen in the parking lot spectrum and field of view scaling. The two days analysed match those from the downlink.} \label{fig:uplink_UC}
\end{figure}

\begin{figure}[h!]
\centering
\includegraphics[width = \linewidth]{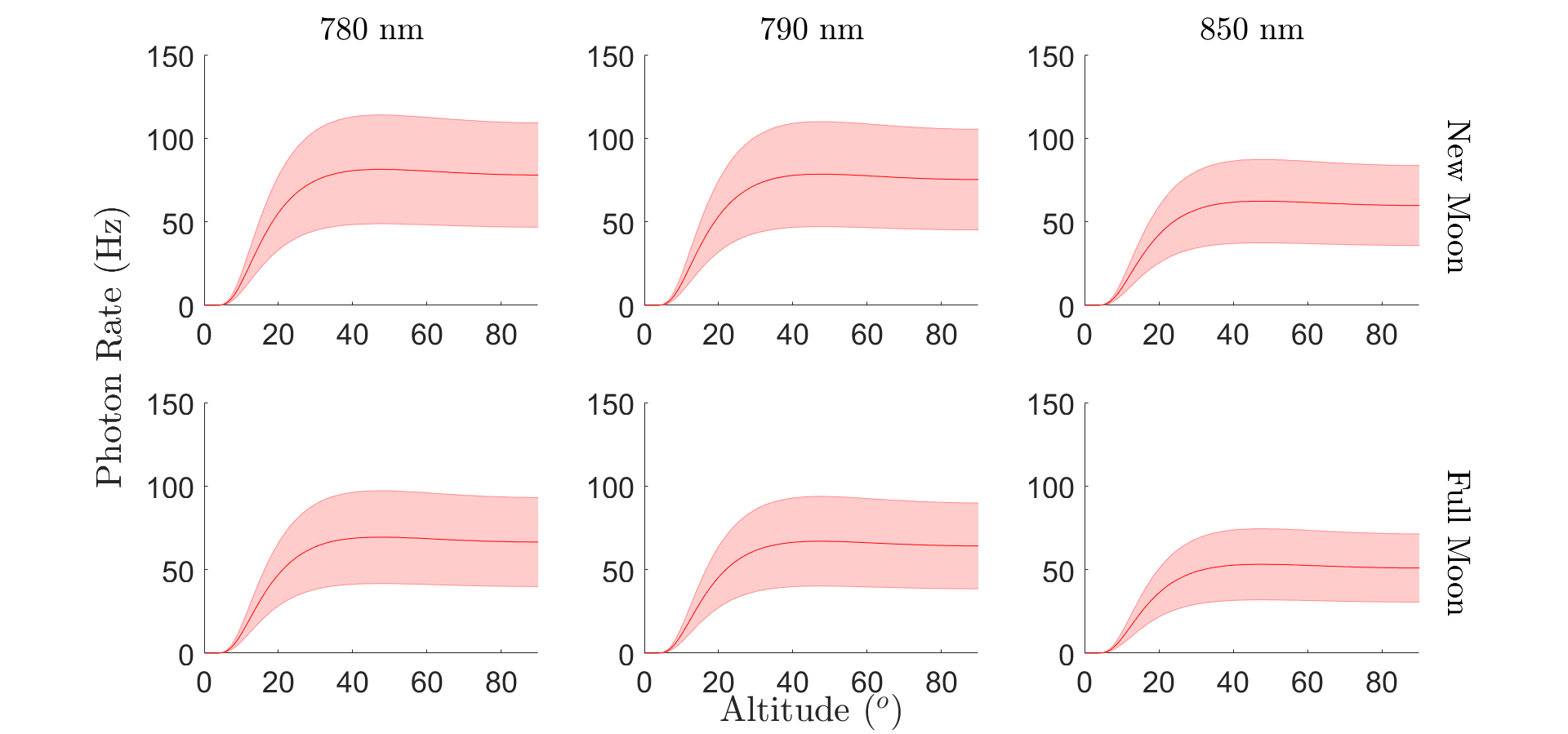}
\footnotesize
\caption{Expected photon counts reaching QEYSSat for an uplink configuration from the QGS-RAO using the VIIRS method. The VIIRS uncertainty is from 40\% variation seen in the parking lot spectrum and field of view scaling. The two days analysed match those from the downlink.} \label{fig:uplink_RAO}
\end{figure}

At QGS-UC, we see a radiance reaching 105 nW/cm$^2$/sr from the VIIRS data with a maximum uncertainty of 15\%. By using the atmospheric transmission model presented in \cite{bourgoin2013comprehensive}, we found that the photon rates in the desired uplink wavelengths, presented in Figure~\ref{fig:uplink_UC} are in the order of 2000\:Hz. In comparison, the QGS-RAO, located 40\:km from the University of Calgary in a dark sky area, has a maximum radiance of 3.4\:nW/cm$^2$/sr\:$(\pm 15\%)$ on the same dates. This gives approximately 100 nW/cm$^2$/sr difference between the urban area where QGS-UC was located and a dark sky area. Figure~\ref{fig:uplink_RAO} shows we can predict a photon rate in the order of a hundreds per second within the QEYSSat spectrum.


\section{Discussion}\label{sec12}

To properly discuss the collected data, it is necessary to regard uplink and the downlink separately. When a ground station is used as the receiver, certain atmospheric effects, such as scattering are less prevalent \cite{Jennewein2014}. This is due to the bulk of the overall path length of the satellite communication is in the vacuum of space. Therefore, scattering only has an effect on approximately the last 10\% of the path.

\begin{figure}[h!]
\centering
\includegraphics[width = \linewidth]{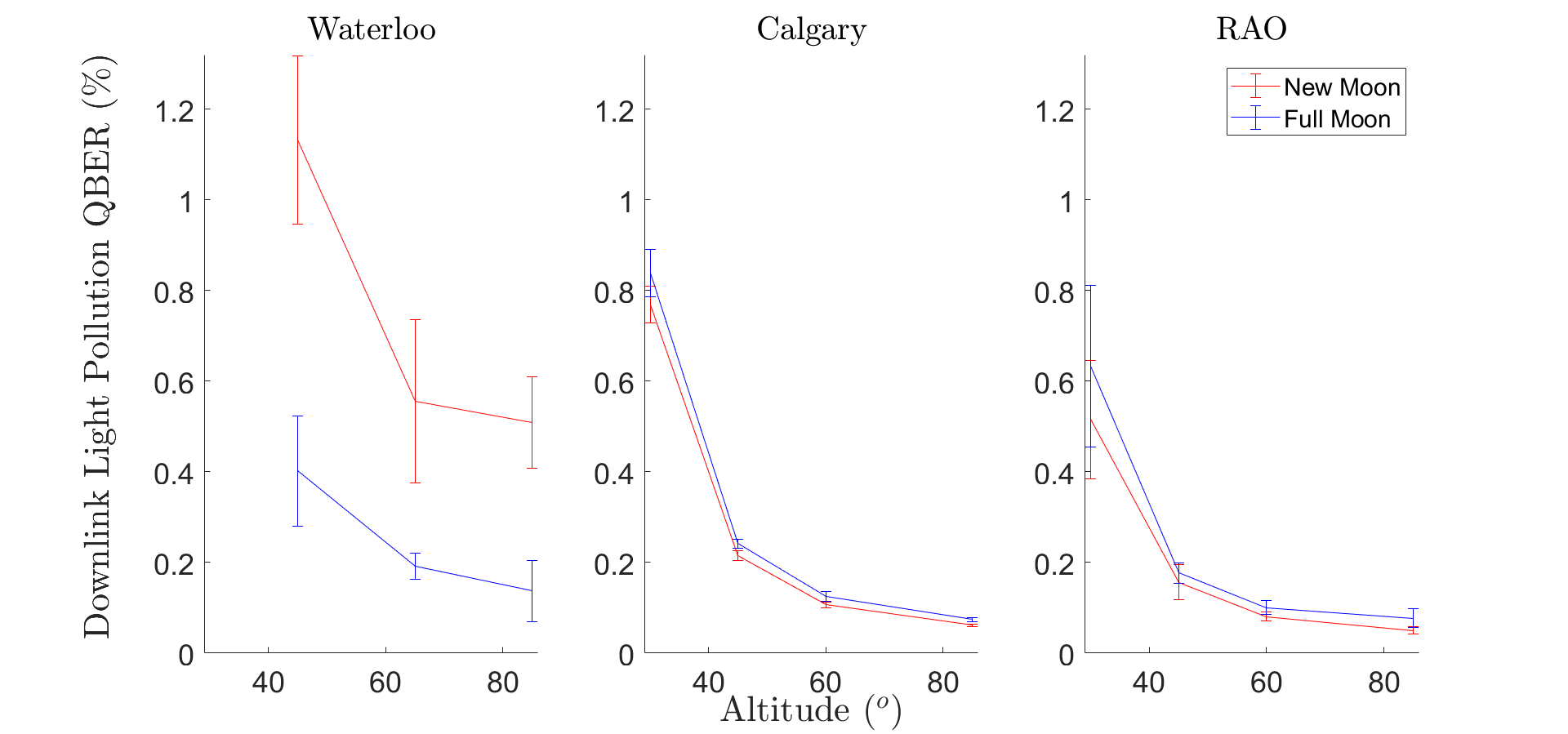}
\footnotesize
\caption{Predicted QBER rates for a quantum downlink pass from QEYSSat's onboard quantum source (850\:nm at 100\:MHz). Downlink azimuthal measurements were averaged together to give the counts at a given altitude; uncertainty comes from the standard deviation of that averaging. For the Waterloo full moon measurement, the data points corresponding to the full moon itself were not included in the averaging. Light pollution is the only source of noise in this QBER calculation. As shown in Table \ref{tab:Locations}, QGS-UW has a wider FOV compared to the telescope at QGS-UC and QGS-RAO sites, and will therefore see more background light pollution.} \label{fig:D_QBER}
\end{figure}

Some considerations need to be made to compare downlink measurements between the QGS-UC/QGS-RAO and QGS-UW sites as the two measurement set ups were very different. The QGS-UW setup was far more mobile, allowing for ease of examination of potential ground station sites. While the telescope systems used at QGS-UC and QGS-RAO work well for analysing current light pollution in the surrounding area but, as QGS-RAO is a fully active ground station (QGS-UC is no longer operational), it is immobile. This lack of mobility puts QGS-RAO at a disadvantage when it comes to seeking out potential ground station locations. However, as QGS-UC and QGS-RAO have successfully performed satellite-to-ground QKD experiments with both the Micius and Jinan-1 satellites, it allows a reasonable benchmark for acceptable levels of light pollution. 

As an additional comment on the measurements and predicted data, we consider the time at which the measurements occurred during each of the studied nights. Each full moon data collection period was designed such that the moon was present during the measurements. While this is an acceptable indicator of the maximum light expected during these periods of moonlight, the measurements taken do not necessarily match the time in which QKD experiments with QEYSSat are expected to occur. QEYSSat, being a sun synchronous polar-orbiting satellite will pass each ground station at close to the same time, between 5:00 UTC and 6:00 UTC, each night. Meaning that during many QKD experiments, the pass will occur before moonrise which will reduce, or even eliminate moonlight.

Other sources of loss including scattering are less prevalent but still need to be analysed while considering the atmospheric conditions at each measurement date.  Humidity coupled with lower temperatures can cause the formation of ice crystals in the air. These ice crystals reflect light in the atmosphere and increase overall sky glow at the NIR spectrum. However, as ice is less transparent to infrared than water vapour, we see a slight reduction to light pollution measurements on colder days \cite{shin2019characteristics}, as shown in Figure~\ref{fig:Downlink} where the average counts are lower in the full moon case. Consequently, the presence of ice crystals lead to an increased QBER for the new moon scenario at QGS-UW in Figure~\ref{fig:D_QBER}.

Considering the measurements shown in Figure~\ref{fig:UofC8Panel}, the largest photon count rate is at 850\:nm in the direction of the horizon. First, we must consider that  the telescope at QGS-UC is optimised for 850\:nm, so the data recorded at 750\:nm and 800\:nm was normalised to facilitate the comparison between the wavelengths measured. The plot in Figure~\ref{fig:UofC8Panel} shows increased photon counts around the horizon angles appearing at first glance to show standard atmospheric sky-glow. Figure~\ref{fig:updown} shows how skyglow is reflected back to the ground station. Indeed, through simple geometry we can determine that the apparent atmospheric thickness is greater toward the horizon producing a greater intensity of sky-glow from a low 30$\degree$ altitude than that of an 80$\degree$ altitude \cite{kaufman1993aerosol}. If we more carefully consider the horizon light observed in the new moon phase, we can contribute it to a combination of several atmospheric lighting effects.

An essential question this study seeks to answer is how much light pollution can be tolerated before the signal-to-noise ratio becomes too low, making satellite quantum communication not possible? There are methods of error correction that can theoretically correct QBER of up to 18\% or more by using two-way classical post processing \cite{gottesman2003proof}. The theoretical limit for the qubit-based BB84 protocol is 11\% \cite{Shor2000}, while the practical limits of QKD with QEYSSat are closer to 5\% \cite{WOS:000365771500011}. For comparison, the  QKD downlink implementation reported with satellite Micius, which used a source emitting  848\:nm signals at 100\:MHz \cite{lu2022micius}, the total photon counts received at the QGS-UC were on the order of $10^4$\:Hz. The noise detections, which include detector dark counts and unwanted light sources, resulted in an error rate of $ < 2\%$.

\begin{figure}[h!]
\centering
\includegraphics[width = \linewidth]{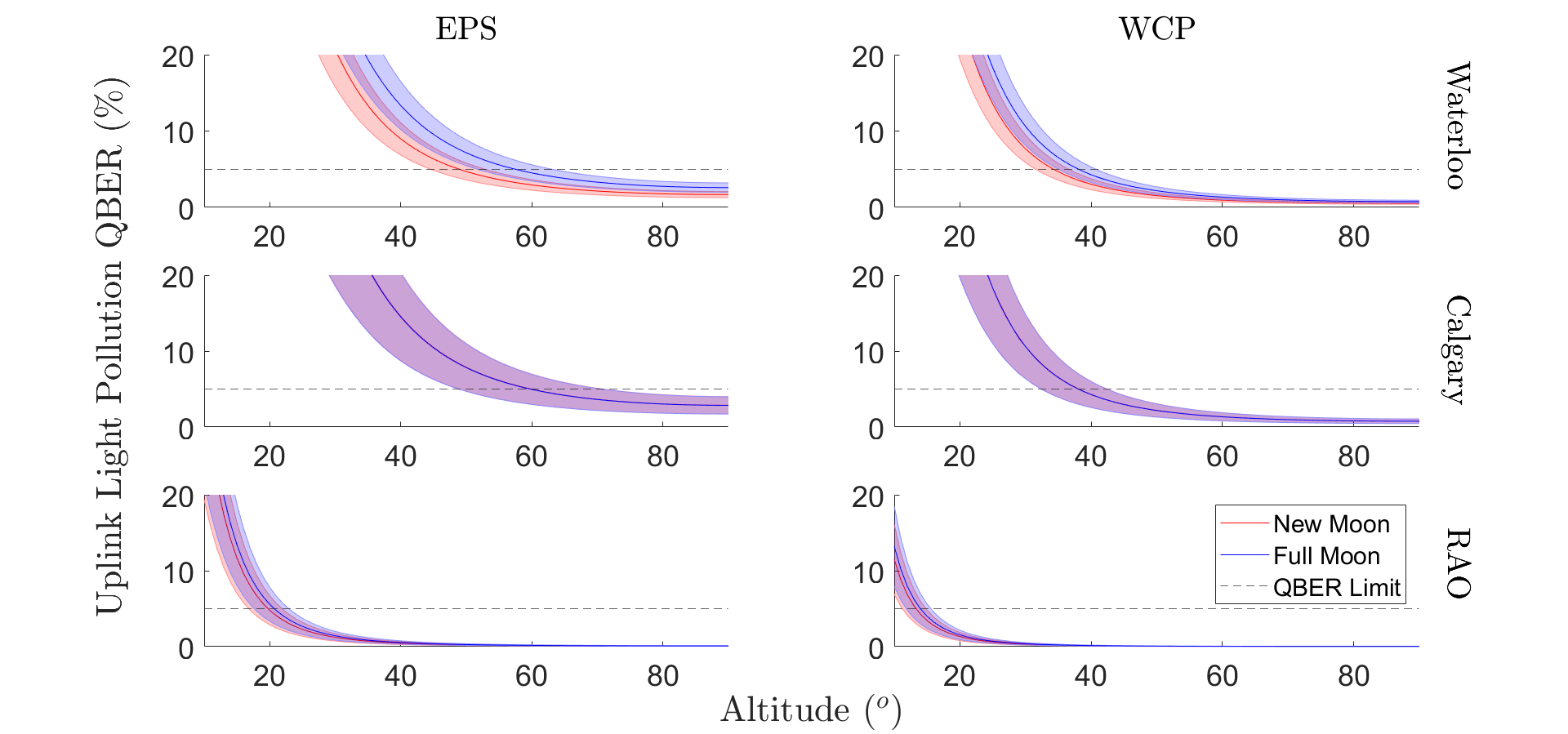}
\footnotesize
\caption{Predicted QBER rates for a quantum uplink pass to QEYSSat using either a weak coherent pulse quantum source (780~nm at 400~MHz) or an entangled photon quantum source (790~nm at 100~MHz). Light pollution estimated from the VIIRS method is the only source of noise in the QBER calculation, with uncertainty  estimation as described in the methods section.} \label{fig:U_QBER}
\end{figure}

Using link-loss models for quantum satellite communication \cite{bourgoin2013comprehensive}, it's possible to derive the expected number of desired photons arriving at a receiver knowing the quantum source rate. Using the light pollution data from Figures~\ref{fig:Downlink} and \ref{fig:UofC8Panel}, the QBER of a downlink can be simulated as shown in Figure~\ref{fig:D_QBER}. As the telescope used at QGS-UC and QGS-RAO is optimized to act as a receiver, it has excellent predicted QBER performance and the simulated results match the observed error rates. QGS-UW however is primarily designed to be a transmitter, as such does not have as good performance; nevertheless, the results show that it can still perform a downlink pass due to its predicted sub 2\% QBER.

Similar to the downlink case, the expected QBER of an uplink pass was derived using the light pollution data from Figures~\ref{fig:uplink_sum}, \ref{fig:uplink_UC} and \ref{fig:uplink_RAO} , with the results shown in Figure~\ref{fig:U_QBER}. QGS-UC and QGS-UW sites are both urban locations, and as such the light pollution is predicted to make it impossible to have a secure connection until QEYSSat is at least above 40$\degree$ altitude. In contrast, the QGS-RAO is in a rural dark-sky environment, and we predict that a secure connection is potentially possible at as low as 10$\degree$ altitude.

\section{Conclusion}\label{sec13}

We characterized three Canadian QGS locations to determine their suitability for establishing a quantum link with the upcoming QEYSSat mission regarding their local light pollution levels by studying local measurements and satellite-borne instrument data.  We determined that despite the visible sources of ground-based light pollution, the photon rates appear to be in the tolerable range of 1000\:Hz for downlink quantum communication, and several thousand Hz when considering the uplink direction. Our measurements and model predict that at the QGS-UC and QGS-UW sites (urban locations), a secure connection to QEYSSat will not be possible until the satellite is at least above 40$\degree$ altitude due to light pollution. In contrast, the QGS-RAO is in a rural dark-sky environment, and a secure connection is potentially feasible at as low as 10$\degree$ altitude.

 We compared local measurements of photon background counts at QGS sites with the light pollution estimates from the VIIRS data product, and found that these results are in reasonable agreement. Therefore, this method can help estimate an upper bound on the expected background counts at a potential QGS site using the satellite-based measurements. Consequently, we determined that the background photon rates at the three QGS sites should not prevent successful quantum uplink and downlink experiments. Furthermore, we collected data at the dark-sky area QGS-RAO, which showed less than 10\% of the background value of QGS-UC. We must still consider how the proximity of a potential QGS to existing networking infrastructure may limit how far into a dark sky area a QGS can be deployed. 
 
Our method of using satellite-based data (such as VIIRS data product) comes with a large uncertainty, in particular given the local spectra of the background light, however, it allows  to remotely and efficiently estimate the suitability of a ground site. Future studies will be include more detailed local spectral measurements and also assess the impact of daily and seasonal variations of local conditions, such as temperature and humidity. We also plan several extensions of  this approach by, for instance, using ground-based sky monitor systems and possibly  airborne detectors to improve the local measurements used to scale the satellite data. Ultimately, a conclusive validation of the model will be provided by observations and experiments performed with QEYSSat.


\backmatter

\section*{List of Abbreviations}
\begin{itemize}
    \item \textbf{QGS} - Quantum Ground station
    \item \textbf{QEYSSat} - Quantum Ecryption and Science Satellite
    \item \textbf{VIIRS} - Visible Infrared Imaging Radiometer Suite
    \item \textbf{QKD} - Quantum Key Distribution
    \item \textbf{QBER} - Quantum Bit Error Rate
    \item \textbf{QGS-UW} - Quantum Ground Station, University of Waterloo
    \item \textbf{QGS-UC} - Quantum Ground Station, University of Calgary
    \item \textbf{QGS-RAO} - Quantum Ground Station, Rothney Astrophysical Observatory
    \item \textbf{DNB} - Day and Night Band
    \item \textbf{DMSP} - Defense Meteorological Satellite Program
\end{itemize}

\section*{Declarations}
\subsection{Availability of data and materials}
The datasets used and/or analysed during the current study are available from the corresponding author on reasonable request.
\subsection{Competing interests}
The authors declare that they have no competing interests.
\subsection{Funding}
NSERC, Alberta Ministry of Technology and Innovation's Major Innovation Fund for Quantum Technologies, CFI, Ontario Research Fund, NSERC Alliance UK-Canada project 'REFQ', and NSERC Alliance Quantum Consortium 'QUINT',  NRC HTSN-343 project.
\subsection{Authors' contributions}
Mathew Yastremski: Manuscript writing, data acquisition and analysis, figures and plots.
Paul J. Godin: Manuscript writing, data acquisition and analysis, figures and plots.
Nouralhoda Bayat: Data acquisition and analysis. 
Sungeun Oh: Link loss modeling, data acquisition and analysis, figures and plots. 
Ziheng Chang: Data acquisition.
Katanya B. Kuntz: Manuscript writing and analysis.
Daniel Oblak: Manuscript writing and analysis.
Thomas Jennewein: Manuscript writing and analysis.
All contributed to the discussions and reviews.
\subsection{Acknowledgments}
The authors would like to thank Erhan Saglamyurek, Karabee Batta, Quinn Rupert, Abhijeet Alase, Brian Moffat, Veronica Chatrath, and Guanlun Zhao for their assistance with this project.

The VPN46A1 data product was acquired from the VIIRS (Visible Infrared Imaging Radiometer Suite) DNB (Day/Night Band) Level-1 and Atmosphere Archive and Distribution System (LAADS) Distributed Active Archive Center (DAAC), located in the Goddard Space Flight Center in Greenbelt, Maryland (https://ladsweb.nascom.nasa.gov/).

%

\end{document}